\documentclass{IEEEtran}

\usepackage{graphicx}
\usepackage{xcolor}

\begin{document}


\title{Fiduciary Responsibility:  Facilitating Public Trust in Automated Decision Making}
\author{Shannon B. Harper and Eric S. Weber}


\maketitle

\begin{abstract}
Automated decision-making systems are being increasingly deployed and affect the public in a multitude of positive and negative ways. Governmental and private institutions use these systems to process information according to certain human-devised rules in order to address social problems or organizational challenges. Both research and real-world experience indicate that the public lacks trust in automated decision-making systems and the institutions that deploy them. The recreancy theorem argues that the public is more likely to trust and support decisions made or influenced by automated decision-making systems if the institutions that administer them meet their fiduciary responsibility. However, often the public is never informed of how these systems operate and resultant institutional decisions are made. A ``black box'' effect of automated decision-making systems reduces the public's perceptions of integrity and trustworthiness. Consequently, the institutions administering these systems are less able to assess whether the decisions are just.   The result is that the public loses the capacity to identify, challenge, and rectify unfairness or the costs associated with the loss of public goods or benefits. 

The current position paper defines and explains the role of fiduciary responsibility within an automated decision-making system.  We formulate an automated decision-making system as a data science lifecycle (DSL) and examine the implications of fiduciary responsibility within the context of the DSL.  Fiduciary responsibility within DSLs provides a methodology for addressing the public's lack of trust in automated decision-making systems and the institutions that employ them to make decisions affecting the public. We posit that fiduciary responsibility manifests in several contexts of a DSL, each of which requires its own mitigation of sources of mistrust. To instantiate fiduciary responsibility, a Los Angeles Police Department (LAPD) predictive policing case study is examined.  We examine the development and deployment by the LAPD of predictive policing technology and identify several ways in which the LAPD failed to meet its fiduciary responsibility.
\end{abstract}





\section{Introduction}

Automated decision-making systems are being rapidly deployed in the United States and internationally and affect the public in a multitude of positive and negative ways. Private and governmental institutions (i.e. societal institutions) use these systems to process information according to certain human-devised rules in order to address social problems or organizational challenges. These systems are often created using mathematical formulas or algorithms that are processed through computers to find commonalities among large datasets. For example, police departments have designed (with the assistance of data scientists) predictive policing algorithms to analyze massive amounts of pre-existing crime data to identify communities that have a high risk of crime; or past arrests or victimization data to identify individuals/groups who are likely to commit a crime or become a victim. 

Some research suggests that the public lacks trust in automated decision-making systems and the institutions that deploy them \cite{sapp2021public,kao2022effect}. The recreancy theorem \cite{sapp2009consumer} argues that individuals are more likely to trust and support decisions influenced by automated decision-making systems if the institutions that administer them behave with integrity (i.e. fiduciary responsibility) and competency. However, often, the public is never informed of how these systems operate and resultant institutional decisions are made. A “black box” effect reduces the public’s perceptions of automated decision systems’ integrity and trustworthiness. Consequently, the institutions administering these systems are less able to assess whether the decisions suggested are just; and the public loses the capacity to identify and challenge unfairness, or the costs associated with the loss of public goods or benefits. 

The current position paper examines fiduciary responsibility \cite{sapp2021public,sapp2009consumer} within the context of a data science lifecycle (DSL). There are many DSLs that affect individuals and the public at large, thus requiring institutional fiduciary responsibility. Examples of these DSLs include predictive policing \cite{mohler2015randomized,lum2016predict,ensign2018runaway}, application processing (e.g. loans, school admissions, etc), autonomous vehicles \cite{hengstler2016applied} and robotics \cite{siau2018building}, and government network surveillance and national security \cite{sapp2021public}. DSLs provide a holistic framework for describing processes and attributes of automated decision-making systems. A DSL has three layers: a (1) pre-processing layer, a (2) model building layer, and a (3) post-processing layer (see Subsection \ref{ssec:recreancy}). Drawing from the recreancy theorem in quantifying the public's trust in automated decision-making systems, the current paper focuses on fiduciary responsibility within the third layer of the DSL.  There is already a significant body of work to substantiate fiduciary responsibility within the early layers of DSLs (see Subsection \ref{ssec:related}).  Our contribution is two-fold: (i) to analyze the notion of fiduciary responsibility within the third layer of a DSL, and (ii) assert that reducing the black box effect in that layer is necessary for institutions to meet their fiduciary responsibility (see Section \ref{sec:dsl}).  We discuss the role of fiduciary responsibility within DSLs, which provides a methodology for addressing the public’s lack of trust in automated systems and the institutions that employ them to make decisions affecting the public (see Section \ref{ssec:fr-in-dsl}). We posit that fiduciary responsibility appears in several contexts of a DSL, each of which requires its own mitigation of sources of mistrust.  To instantiate our view of fiduciary responsibility within a DSL, a Los Angeles Police Department (LAPD) predictive policing case study is examined (see Section \ref{sec:case}).  We examine the development and deployment by the LAPD of predictive policing technology and identify several ways in which the LAPD failed to meet its fiduciary responsibility.  We further discuss actions and mechanisms which the LAPD could have utilized in an effort to meet its fiduciary responsibility.

The current position paper is situated in the relevant sociological literature concerning public trust in technological innovations. It provides a novel and potentially impactful framework to address and facilitate fairness, accountability, and transparency in automated decision-making systems, which spans the DSL workflow. Our analysis has a specific focus on building trust in the post-processing layer/stages. We also build on prior work to demonstrate how bias can manifest in the data acquisition, model building, and post-processing DSL layers/stages, requiring distinct mitigation strategies.

\subsection{Important Terminology}

We will be using several phrases to describe, more or less, the same phenomenon that affects the public. These phrases are: (1) automated decision-making systems, (2) artificial intelligence (AI), and (3) data science lifecycles. We will be using them interchangeably, dependent upon context, though we acknowledge that they are not identical. ``Automated decision-making systems'' is a common phrase used in sociology literature \cite{dobbe2018broader,lee2004trust,JS16,MATWF,BS15} (but not exclusively \cite{mouzannar2019fair}) and refers to institutional implementation of a mechanism--often without a human-in-the-loop--for making a decision and subsequently deploying an action that has an appreciable effect upon an individual or community. From our view, this is the best description of the systems we consider here when viewed from the perspective of the public or stakeholders. As we will describe in Section \ref{sec:dsl}, when the DSL occurs within a black box from the public's perspective, it is acting as an automated decision-making system. Artificial intelligence, for our purposes, is a methodology that (in part) mechanizes the automated decision-making system or appears as a stage within a DSL. As such, the AI moniker is narrower in scope than the overall pipeline that we have in mind for examining fiduciary responsibility. However, as it is very common in the literature, we still use it, particularly when we are referring to the work of other researchers. As emphasized in \cite{knowles2021sanction}, AI also often operates within a black box from the perspective of the public. Finally, we use the phrase ``data science lifecycle.'' ``Data science'' refers to methods and algorithms that interact with data, typically through acquisition, management, analysis, modeling, and reasoning \cite{biswas2021art,nguyen2019machine,olson2016evaluation}. As such, it encompasses more than statistics or data mining \cite{wickham2019data}. The term ``lifecycle'', or ``pipeline'', is becoming more common in the literature \cite{hong2017build,todd2017computing,garcia2018context,polyzotis2018data,zhou2019how,wing2019data,ashmore2021assuring,biswas2021art}. We describe our usage of the phrase in Section \ref{sec:dsl}. To emphasize the point here, DSL is meant to be an encompassing term that includes both AI and automated decision-making systems. Ultimately, using the notion of fiduciary responsibility, we will demonstrate that to facilitate public trust in these systems, much of the DSL should operate in view of stakeholders.

\section{Conceptual Framework: Fiduciary Responsibility} \label{sec:fr}



\subsection{The Recreancy Theorem and Fiduciary Responsibility} \label{ssec:recreancy}

As conceptualized by Sapp et al. \cite{sapp2009consumer}, the recreancy theorem argues that the public’s trust in public and private societal institutions is explained by their perceptions of the institution's competence (i.e. skill, ability, experience), and their confidence that the institution will behave with integrity (i.e. honesty, ethical standards), also known as fiduciary responsibility \cite{freudenburg1993risk}. Benevolence is a third central component of public trust, which involves the perceived extent to which the institution is concerned about citizens' welfare. Multiple scholars have argued that when societal institutions (with a wide spectrum of roles and responsibilities) fail to build trust among the populations they serve (i.e. reflect recreancy), society’s ability to function is detrimentally effected \cite{alario2003paradoxes, freudenburg1993risk,roth1978economy,sapp2021public}. Some have argued that trustworthiness rather than trust signifies public opinions of societal institutions' behaviors. \cite{colquitt2007trust,ball2019institutional,sapp2021public}.

Sapp et al. \cite{sapp2021public} (see also \cite{freudenburg1993risk}) define trustworthiness as institutional behaviors that give citizens reason to have confidence in their performance. Interpersonal trust is embedded in the recreancy theorem wherein there is a perceived connection between a societal institution and individuals in the population they serve \cite{sapp2021public}. When the public perceives an institution as meeting their expectations of competent, honorable, and benevolent performance, interpersonal trust between the two is established \cite{barber1983logic,blomqvist2005trust,blomqvist2005playing,deutsch1958trust}, and recent research affirms this contention  \cite{sapp-dm2009consumer,ball2019institutional,kim2012participation,sapp2021public}. Such perceptions influence whether the public trusts and supports technological innovations (such as automated decision-making systems) proposed or administered by these institutions \cite{barber1983logic,blomqvist2005trust,hardin2001conceptions,earle1995social,siegrist2000perception}. Additionally, recent research reveals that the public is more likely to trust social institutions when their newly administered AI does not negatively affect social justice--i.e. protects the interests of vulnerable/marginalized populations \cite{cooper2008importance,kim2012participation,sapp2021public,schoorman2007integrative}.

The degree to which the public trusts institutions in their administration of automated decision-making systems often influences whether those systems are abandoned or used long-term \cite{sapp2009consumer,sapp2021public}.
Fiduciary responsibility \cite{sapp2021public} is an integral tenet of the recreancy theorem and refers to public perceptions of the integrity demonstrated by societal institutions, which influences individuals’ trust in and support for automated decision-making systems administered by those institutions. In the current paper, we specifically examine fiduciary responsibility in societal institutions’ development and administration of automated decision-making systems because embedded processes of data collection, data modeling, and prediction output influence whether the public will perceive those institutions as having integrity. Following automated predictions, some action typically occurs. For example, in a system that processes loan applications, the system decides whether to approve a loan application after making a prediction on whether an applicant will likely repay the loan. The prediction, and subsequent decision, is based on a data collection method and a model developed by the designer of the system. Such automated decision-making systems are part of a larger group of data driven processes that are often called DSLs or sometimes, ''data science pipelines''. Societal institutions that administer algorithm-driven automated decision-making systems with honor--e.g. concern for and attention toward minimizing bias/racism and privacy infringement--are more likely to be trusted by the public \cite{sapp2021public}. We will show that automated decision-making systems that operate within a ``black box'' (where data scientists and institutional staff make system development and administration decisions ``in the dark'') absent public technological knowledge, informational awareness, and scrutiny, hinders an institution's efforts to meet fiduciary responsibility, and consequently establish trust.

\subsection{Fiduciary Responsibility and Public Trust in Technical Innovations} \label{ssec:innovations}

The recreancy theorem, as applied specifically to technical innovations, delineates three dimensions of public acceptance of technical innovations. As stated in \cite{sapp2021public}, the recreancy theorem:
\begin{quote}
complements technology adoption models in that it focuses upon public assessments of innovations as they are managed by societal institutions, thereby providing conceptual congruity between technology adoption and public assessments of institutional competency and integrity. 
\end{quote}

To that end, \cite{sapp2009consumer,sapp2013science} argue that the public is more likely to trust and accept the implementation and use of technologies that pose a risk to their welfare or interests provided they meet the following requirements. First, they (the technologies and/or institutions deploying those technologies) must possess \textbf{technical competency}, meaning that they must have the capability to perform the analysis and/or actions required for the intended purpose of the technologies. Second, they must have a \textbf{public benefit}, meaning that these technologies should improve upon a specific problem of public concern upon deployment. And third, they must meet \textbf{fiduciary responsibility}, meaning that the algorithms are designed and employed with integrity, and are in fact performing in the way that they are intended, without disparate impacts or misuse by agents.  

Each of these dimensions warrants investigation with respect to the public's trust in AI deployment in particular. We describe in Subsection \ref{ssec:related} prior works that have related themes, especially methods for formalizing, analyzing, and quantifying technical competency. As such, we will not address that dimension here, nor will we consider the public benefit aspect of the recreancy theorem. We will focus on fiduciary responsibility, building upon recent research findings that fiduciary responsibility is of particular importance among the public, especially for the protection of vulnerable populations. Scholarship suggests that institutional administration of technological innovations raises multiple concerns specific to integrity, including invasion of privacy, loss of personal health data, and unfair monitoring or targeting of particular individuals or groups, which may sometimes involve racial bias \cite{anderson2018artificial, hitlin2019facebook, olmstead2017americans, rainie2017code, rainie2017internet, sapp2021public}.


Further, \cite{sapp2021public} discusses a conceptual understanding of public opinions of network surveillance and empirically documents public demand for network surveillance that fosters goals of social justice more so than goals of self-interest. The findings are based on a nationwide survey of adults concerning governments' use of network surveillance. Additionally, the utilization of a technological innovation is often perceived as socially just when it protects the rights of marginalized/vulnerable populations such as people of color, women, and LGBTQ+ individuals \cite{sapp2021public}.


The recreancy theorem asserts that the public's acceptance of technical innovations depends on the public's perception that institutions fulfill fiduciary responsibility while deploying those technologies. Therefore, it is incumbent upon institutions to ensure that the public is confident that fiduciary responsibility is met by the technologies in use.  By utilizing the formalism of a DSL, we will argue that this can best be done through transparency at multiple stages of the lifecycle, especially those stages that occur in the third layer identified as interpretation and communication (see Figures \ref{fig:1-blackbox} and \ref{fig:2-blackbox}).



\subsection{Related Work} \label{ssec:related}

Public trust in automated decision making systems or artificial intelligence is, \emph{prima facie}, both important to establish and difficult to formalize. For example, scholars \cite{lewis1985trust,hofstede2006intrinsic,WHKA14} have put forward competing formal definitions of trust, either interpersonal or institutional, and these definitions have various dimensions. Several high profile institutions, \cite{ibm2020trusting,european2020white} just to name a few, acknowledge the importance of establishing trust in AI while using the term in a colloquial sense. Early efforts to formalize trust in various forms of digital interactions appear in \cite{YH08,harper2014trust}.


Recent work has endeavored to describe trust in AI in several different ways on a technical basis. Several authors have posited a formal definition of trust in AI by drawing on prior work in formalizing interpersonal and institutional trust \cite{jobin2019global,toreini2020relationship}. Other authors have proposed methods for quantifying, or establishing, trust through established legal or public policy structures \cite{arnold2019factsheets,knowles2021sanction}.  Still, others have distinguished between trust in AI versus trustworthy AI \cite{jacovi2021formalizing}.  There are multiple ways in which our use of the recreancy theorem in general, and fiduciary responsibility in particular, in the context of public trust in AI are related to these other works.  We describe those relations here.

It is well understood that public trust in AI has multiple dimensions.  The three facets of the recreancy theorem--competency, benefit, and fiduciary responsibility--are reflected in others' investigations into the question of how AI can be trustworthy.  Indeed, \cite{jobin2019global} finds similar dimensions as the recreancy theorem:
\begin{quote}
To investigate whether a global agreement on these questions is emerging, we mapped and analyzed the current corpus of principles and guidelines on ethical AI. Our results reveal a global convergence emerging around five ethical principles (transparency, justice and fairness, non-maleficence, responsibility and privacy), with substantive divergence in relation to how these principles are interpreted.
\end{quote}
Toreini et al. \cite{toreini2020relationship} describe trust in AI as distinct from trustworthy ML by utilizing the ABI framework--Ability, Benevolence, and Integrity--posited by \cite{mayer1995integrative} to model organizational trust.  We note that Mayer's ABI framework and the recreancy theorem run parallel to each other in their dimensions, while Toreini and colleagues' notion of trustworthy as a part of trust reflect the technical competency of the recreancy theorem.  This is borne out by several related notions such as trustworthy AI and human-AI trust.

In general, \emph{trustworthy AI} refers to the competence of the AI algorithm, with competence defined in terms of a contract.  This is precisely how Jacovi et. al. \cite{jacovi2021formalizing} define the notion: ``an AI model is trustworthy to some contract if it is capable of maintaining this contract.''  Similarly, \cite{hawley2014trust} describes trust facilitated through a  commitment (i.e. a contract). This commitment also involves expectations by the trustor about the competence and willingness of the trustee.  Knowles et. al. \cite{knowles2021sanction} developed a foundation for trust in AI through a transparent and understandable regulatory system.  
They argue that a regulatory system can potentially bridge the gap between trustworthy and trust.  Such a regulatory system can be a mechanism for extending the public's trust from acceptance that a particular AI meets the required technical competence to meeting the required fiduciary responsibility as well.

Jacovi et. al. \cite{jacovi2021formalizing} further propose a formal definition of \emph{human-AI} trust via adapting the sociological definition of interpersonal trust. The definition has several dimensions, including trustworthiness of the algorithm and warranted trust possessed by the human who is at risk of the AI's actions. They posit that an algorithm is trustworthy if it can uphold a contract, and a human possesses warranted trust of AI if the human has reason to accurately anticipate the impact of the AI's decisions.  The notion of trustworthy put forth in \cite{jacovi2021formalizing} need not require an individual to understand the inner-workings of the algorithm.

Others have argued for facilitating public trust in AI through an understanding of the workings of the algorithms, either by individuals that are subject to AI or by experts who can vouch for the competency of the algorithms. A large body of literature discusses Explainable AI (XAI) as one method for ensuring trust between users and AI \cite{ribeiro2016should,mittelstadt2019explaining}. Documentation, such as AI factsheets \cite{arnold2019factsheets,richards2020methodology} (for example the European requirements for factsheets \cite{european2020white}) or declarations of conformity \cite{hind2018increasing}, is a potential way to build public trust by facilitating auditing of AI.  Both of these mechanisms are designed to meet the perception of the algorithms' technical competency among the public, and consequently establish trust.

Our discussion here concerns (predominantly) the public's trust in technological innovations as stakeholders within the environment that the innovation is deployed.  As such, the recreancy theorem is one of multiple technology adoption models (TAM).  A TAM that centers on user acceptance was introduced by Davis \cite{davis1989perceived,davis1993user}.  In our analysis, the users are the institutions deploying the technology, and thus the recreancy theorem is the relevant model for understanding public acceptance rather than user acceptance.

\section{Public Trust in Data Science Lifecycles} \label{sec:dsl}

We seek to formalize the notion of fiduciary responsibility within a DSL as a way to facilitate public trust in systems that utilize AI. We will adapt the formal description of a DSL as discussed in \cite{biswas2021art}, which decomposes a DSL into three layers and each layer into multiple stages. We will argue that formalizing a common specification of a DSL can facilitate multiple mechanisms through which to establish public trust. Finally, we will show how the requirements of fiduciary responsibility are embedded within a DSL, and how the common specification of a DSL can facilitate the meeting of those requirements by organizations and institutions that design and deploy them.

\subsection{DSL Formalism}

Data driven processes follow varied forms and take on many functions, but recent work has been made to accurately describe these processes in ways that facilitate our discussion of fiduciary responsibility. A recent study \cite{biswas2021art} was conducted that comprehensively described and classified DSLs. The study found that some DSLs have very few stages, as few as 4 or 5, while others have as many as 11 (as suggested by \cite{ashmore2021assuring}). Many of the publicly available DSLs identified in the study were used only for academic or competitive (e.g. Kaggle) purposes and were not in fact deployed in a real-world environment. These DSLs typically lacked the stages that form our focus, i.e. those that occur after the prediction stage within the DSL.

\begin{figure*}[!t] 
\centering
\includegraphics[width=14cm]{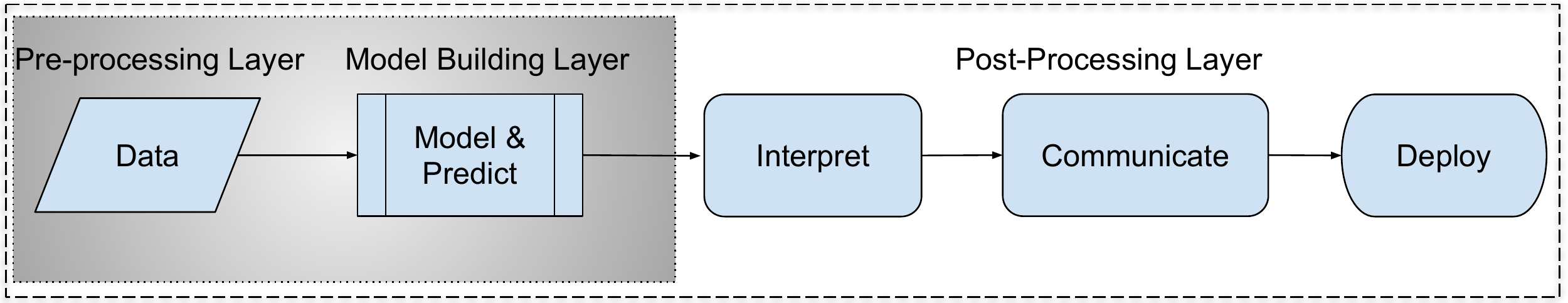}
\caption{A reduced model of a DSL. The pre-processing and model building layers operate within a black box from the perspective of the constituent/stakeholder, while the post-processing layer operates openly from the perspective of the constituent/stakeholder.}
\label{fig:1-blackbox}
\vspace{5mm}
%
\includegraphics[width=14cm]{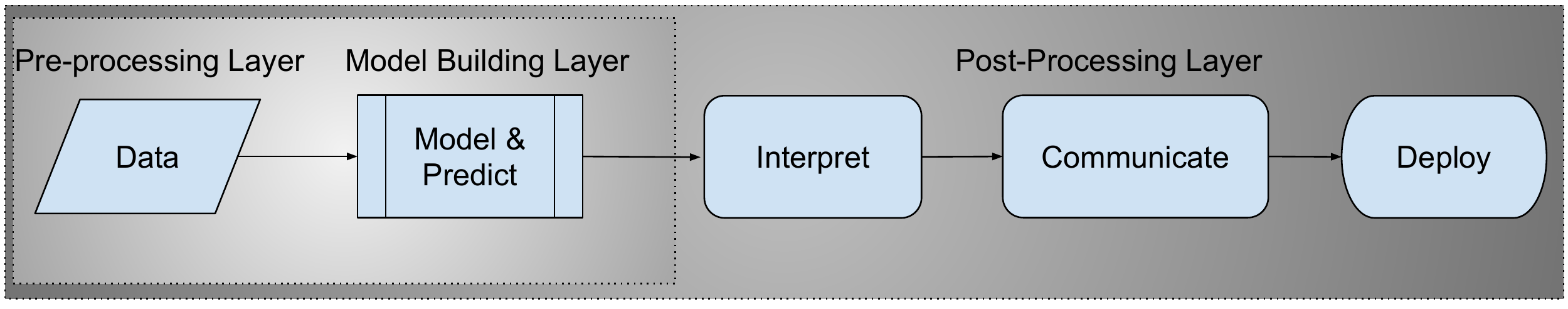}
\caption{A reduced model of a DSL. Here, all three layers operate within a black box from the perspective of the constituent/stakeholder, while the post-processing layer operates openly from the perspective of the agents utilizing the DSL.}
\label{fig:2-blackbox}
\end{figure*}

Following the specification from \cite{biswas2021art}, a DSL has three layers: a pre-processing layer, a modeling building layer, and a post-processing layer. The pre-processing layer consists of three stages: data acquisition, data preparation, and data storage. The model building layer consists of five stages: feature engineering, modeling, training, evaluation, and prediction. Finally, the post-processing layer consists of interpretation, communication, and deployment. The DSLs that we consider should explicitly have all of the 11 stages laid out in \cite{ashmore2021assuring,biswas2021art} (see Subsection \ref{ssec:fr-in-dsl}), but our focus will be on the latter stages within the overall process. We therefore depict the DSL (in a reduced form from \cite{biswas2021art}) for our purposes in Figures \ref{fig:1-blackbox} and \ref{fig:2-blackbox}.

The stages within the post-processing layer are described as follows in \cite{biswas2021art}:
\begin{quote}
\textbf{Interpret:} The prediction result might not be enough to make a decision. We often need...post-processing to translate predictions into knowledge. \smallskip

\textbf{Communicate:} ...we might need to communicate with the involved parties (e.g. devices, persons, systems) to share and accumulate information. \smallskip

\textbf{Deploy:} The built DS solution is installed in its problem domain to serve the application... \smallskip
\end{quote}
These descriptions of the stages within the post-processing layer, as well as those of the  stages in the other layers, refer largely to DSLs that are fully automated. However, in the contexts that we are considering--predictive policing, application processing, etc.--there often is a human-in-the-loop (as described in the Case Study presented in Section \ref{ssec:black}). Specific to the LAPD case study we describe in Section \ref{sec:case}, hot spots predictive policing uses an algorithm in the pre-processing layer that attempts to predict high crime areas. Next, police personnel interpret the prediction to deploy crime-reducing resources, e.g. an increased number of foot-patrols in that area. As a result, our view of the post-processing layer is distinct from, though still related to, the view in \cite{biswas2021art}. For our purposes, we describe the stages of the post-processing layer as follows. 
\begin{quote}
\textbf{Interpret:} The prediction requires evaluation by an agent in order to make a decision or recommendation. The evaluation may involve the context in which the prediction occurs and/or additional information that is not available to the model building layer of the DSL. Here, the agent can be a cyber or physical system, but is likely to be human.  \smallskip

\textbf{Communicate:} The agent(s) that evaluate the prediction communicate the interpretation to other agents for the purposes of making a decision. These other agents can also be cyber, physical, or human.  \smallskip

\textbf{Deploy:} Based on the prediction and interpretation, a decision is made. Subsequently, an action is taken; usually this action is taken in the physical world and may involve the deployment of resources. \smallskip
\end{quote}

Returning to our predictive policing example, we note that the public perceives all layers and stages of the DSL occurring within a black box--this occurred specifically in the implementation by the LAPD, as we document in Section \ref{sec:case}. Most notably, the post-processing layer, which operates with humans-in-the-loop, is obscured from public input, scrutiny, or accountability. 

\subsection{Formalizing DSLs to Facilitate Public Trust}

There are multiple mechanisms for facilitating public trust in AI through formalizing DSLs. We will see shortly how the public’s perception of institutions’ fiduciary responsibility can be promoted by following a formal DSL framework. Independent of our thesis regarding fiduciary responsibility, the DSL framework can assist in establishing public trust in
the technologies that are deployed.

Establishing a common DSL format can facilitate consistent development and maintenance, as well as the production of useful DSL documentation when considered as a software engineering endeavor \cite{biswas2021art}. In turn, the common DSL format and concomitant documentation can facilitate trust in AI-as-an-institution, as argued in \cite{knowles2021sanction}. The DSL framework provides multiple specifications that can be utilized for developing regulatory infrastructure, contract formation, fact sheets, and other ``structural assurances'' for facilitating the trustworthiness of AI deployment.

In addition, a common DSL format can also reduce the black box effect, akin to the assertion by \cite{wachter2017counterfactual}, through providing a description of the internal workings of the black box in order to rectify the knowledge gap. The common DSL format decomposes a larger black box into smaller black boxes (as depicted in Figures \ref{fig:1-blackbox} and \ref{fig:2-blackbox}), some of which can be open to the public, and others which can be subject to expert auditing through declarations of conformity. While the public may not understand the black boxes, or even the DSL framework proposed in \cite{biswas2021art}, the commonality can help the public conceptualize the documentation requirements and accountability to which DSLs are subject.

We have mentioned just a few elementary ways in which a formal DSL framework can facilitate public trust in AI. We have not fully explored this consideration and there is much more work to do in this regard, but such work is outside the scope of the present paper. Scholars such as Sapp et al. \cite{sapp2021public} have used structural equation modeling to quantify public trust, finding that considerations of technical efficacy and social justice are significantly and equally associated with public trust in and support for government-administered network surveillance (see also \cite{sapp2013science}).


\subsection{Fiduciary Responsibility within Data Science Lifecycles} \label{ssec:fr-in-dsl}

We argue that meeting fiduciary responsibility within a DSL requires the post-processing layer to operate in an open box capacity. The purpose of opening the post-processing layer is to provide stakeholders a context for understanding decisions made, which may build public trust in the action(s) being deployed/implemented. An additional potential benefit of the open box operation is that the public has a means to hold institutions accountable for the decisions made by DSLs. We emphasize here that the open box operation is a necessary but not sufficient condition for meeting fiduciary responsibility.  

As described in \cite{biswas2021art}, some DSLs can be fully automated, such as bank loan application processing systems. In such a fully automated system, from the perspective of the person subject to the decision of that system, it operates entirely within a black box (as depicted in Figure \ref{fig:2-blackbox}). Naturally, the person is likely not provided any information or context regarding the decision and therefore cannot understand the decision process. Absent other mechanisms--such as documentation or auditing by trusted experts--fiduciary responsibility cannot be met in this regime, which in turn precludes the establishment of trust by those persons subject to the DSL.

We consider here DSLs that are not fully automated, but involve humans-in-the-loop, particularly those DSLs in which humans appear in the post-processing layer. For example, we re-imagine the loan application processing DSL in which the prediction of the model is given to a (human) banking specialist. This person will interpret the prediction, perhaps utilizing additional information not available to the model or placing the prediction within a larger context. The specialist then informs (communicates) the applicant of the decision, at which time an action (deployment) is taken, e.g. the loan is fulfilled, or the application is closed. 

We explore the multiple facets of fiduciary responsibility with the DSLs we have just described, focusing on the instantiation of fiduciary responsibility within the post-processing layer. Before doing so, however, we want to acknowledge that fiduciary responsibility manifests within \emph{all} layers and stages of DSLs. Much work has been done already to substantiate fiduciary responsibility within the pre-processing and model building layers of DSLs as we described in Subsection \ref{ssec:related}, and yet more work remains.

For the sake of context, we mention several facets of fiduciary responsibility that appear within the early layers of DSLs, while emphasizing that our comments here are not exhaustive. At the pre-processing layer, individuals who are subject to a DSL expect that data associated to them is accurate and will be kept private. It is likely that informed consent is required at this stage. Individuals also expect that data associated to them and others collectively do not contain bias, or put them at higher risk for adverse decisions or actions. As part of the modeling building layer, individuals expect that the model accurately analyzes the data without introducing spurious effects or amplifying bias \cite{dobbe2018broader}. In addition, at the prediction stage, individuals expect to be able to anticipate the impact of the model's prediction \cite{jacovi2021formalizing} and subsequent decision. Moreover, the individual expects that the DSL is not being misused or abused by the institution or its agents.

Let us now turn to the several stages within the post-processing layer. Again, we are considering DSLs for which these stages are an integral part, since a decision and subsequent action are necessary for the public to be affected by the DSL, independent of whether they are aware of them. Our examples of DSLs above indeed incorporate these stages in some form or other.

The first stage of the post-processing layer is ``Interpret'', by which we mean an actor interprets the prediction made by the model. As mentioned previously, this actor could be cyber or human, though we focus on a human agent here. Fiduciary responsibility requires the agent to make an interpretation which is honest, ethical, and just. The interpretation should reflect the institution's values, mission and goals, as well as uphold the rights and interests of stakeholders and those impacted by the institution's operations. The interpretation should be understandable to those who are subject to the decision of the DSL, and, as we shall discuss shortly, amenable to communication to individuals, stakeholders, auditors, etc. Interpretations should be well-documented and archived. As a reflection of these criteria, an interpretation serves dual purposes.  

The initial purpose of the interpretation is to provide the decision-maker with a fuller understanding of the prediction. For example, from a data science perspective, the prediction may have an associated confidence level or may indicate the most relevant features of the input data leading to the prediction. The interpreter can provide context to these additional pieces of information for the decision-maker: as Dobbe et. al. argue that ``machine learning models should \emph{facilitate rather than replace} the critical eye of the human expert'' \cite[emphasis in the original]{dobbe2018broader}.

The subsequent purpose of the interpretation is to provide 
other entities with a vested interest in the DSL \cite{gilpin2022explanation}--such as the individual (or community) subject to the decisions of the DSL--with the context or the rationale for the decision. We note that the context for the decision-maker and that for the individual are assuredly different, (for example, see \cite{gilpin2022explanation} which identifies function roles that inform the nature of an interpretation or explanation) but likely have much overlap. This is because, in both cases, ultimately a human will want to have some understanding of the model's prediction. An interpretation, while it need not explain the inner workings of the algorithms involved, does have the potential to assure individuals that the algorithms are operating with competency, and is even able to provide individuals a foundation for harmonizing the decision with their anticipation of that decision \cite{jacovi2021formalizing}.

Proper anticipation of the DSL's output is a potential foundation for building trust between an individual (or community) and an institution deploying the DSL \cite{jacovi2021formalizing}.  We have previously described how XAI is one commonly identified methodology for establishing this foundation.  In our DSL, the interpretation can be informed by XAI when utilized within the model-building layer; conversely, a well-formulated interpretation can counteract the lack of explainability when a model does not utilize XAI.  

We further contend that the interpretation of the DSL output should be driven by \emph{values} in addition to technical and explanatory considerations.  Value-laden interpretations can address epistemological issues related to fairness, accountability, and transparency.   As Dobbe et. al. further argue, questions of fairness
\begin{quote}
    illuminate the range of places in the machine learning design process where issues of \emph{epistemology} arise: they require \emph{justification} and often \emph{value judgment} \cite[emphasis in the original]{dobbe2018broader}.  
\end{quote}
Interpretations, which are value-laden, that are embedded as a formal stage of the DSL also situate the causes and effects of the DSL within a broader context of ``the human element'', personalizing both the individual affected by and the institution deploying the DSL.

Our formalism of DSLs, with ``Interpretation'' firmly embedded within the overall process, contextualizes epistemological questions, such as ``Interpretable to whom?'', or ``For what purpose?''  as asserted by Kohli et. al. \cite{kohli2018translation}.  As we have already identified, the interpretation has several potential audiences: (1) anyone else downstream within the DSL, and (2) those affected by the DSL's output.  These audiences then proscribe the purpose(s) of the interpretation: (1) to help actualize the ultimate goal of the DSL, and (2) to assure stakeholders that the DSL's output are properly  anticipated.


The penultimate stage of the DSL is ``Communicate'', yet from the viewpoint of fiduciary responsibility, communicate is the most crucial of all stages. Indeed, as the recreancy theorem measures the public's \emph{perception} of the integrity, honesty, and justness of an institution and its use of DSLs, full and open communication by the institution is of paramount importance for meeting its fiduciary responsibility.  Indeed, communicating the interpretation of the DSL's output to affected individuals or communities actualizes the secondary purpose of the interpretation, thereby establishing a potential foundation for trust.  As observed in our case study on the LAPD's use of a predictive policing DSL in Section \ref{sec:case}, a lack of communication can lead to the public not trusting in the institution's deployment of a DSL.

Communication opens the black box operation of the DSL at least to the extent that, if successful, individuals and/or the public at large can understand the rationale, if not the mechanism, for the decision and subsequent action. This opening of the black box is shown in Figure \ref{fig:1-blackbox}, and is premised on the argument in \cite{knowles2021sanction} that
\begin{quote}
    lack of public trust in AI has little to do with people's inability to understand how AIs work; rather it is a response to an awareness of a lack of structural assurances of the trustworthiness of the AIs pervading society.
\end{quote}
Hence, we envision a DSL with the early layers still operating within a black box, but the post-processing layer operating openly. We argue that this regime of a DSL operation can greatly advance the institution's efforts to meet its fiduciary responsibility as well as provide the potential for establishing ``structural assurances of trustworthiness'' that the public will require for accepting the implementation of a DSL.

The content of the communication by the institution that deploys a DSL consists of multiple aspects. The form of the communication is likely dependent upon the institution and its values, the nature of the decision itself, and the affected party (e.g. whether an individual or a community). We formulate the communication in part around the rhetorical ``Five W's''. The institution informs the affected party of (What:) the decision made and the subsequent action that was/will be taken; (Why:) the interpretation of the prediction within the larger context of the DSL and the institution's mission, the rationale for the decision based on the model's prediction, and the justification for the action based on the prediction and the decision; (How:) what data was used to make the prediction, how the data was utilized, how the input data as well as the model output were interpreted, and the values that informed the interpretation of the model output.

The final stage of the DSL, which we have referred to as ``Deploy'', is the point at which an individual (or a community) is ultimately affected by the DSL. This is not to say that an individual is not ever affected in prior stages--in fact, this is a distinct possibility, e.g. lost privacy--but this stage manifests the ostensible \emph{raison d'etre} of the DSL. We use deploy to be consistent with \cite{biswas2021art}, though our usage is distinct. Here, we think of an institution choosing whether and how to deploy resources--for example, manpower or finances; but at a more basic level, this stage refers to the institution implementing an action.

Fiduciary responsibility requires an institution and its agents to employ honest, ethical, and just actions. In the context of automated decision making, the actions need to be well-founded in the model, prediction, and interpretation, meaning that the actions are justified through accurate models, correct predictions, and valid interpretations. The DSL formalism provides a framework for ensuring that fiduciary responsibility is met at each stage of the lifecycle, particularly through documentation and auditing \cite{knowles2021sanction}.  In addition, the DSL framework instantiates meeting fiduciary responsibility through both technical \cite{ensign2018runaway,jacovi2021formalizing} and ethical \cite{jobin2019global,sapp2021public} dimensions. We note that some of the issues associated to technical and ethical concerns are context dependent.


An institution is required to employ actions that protect the rights of the affected individuals, particularly those of vulnerable populations. In the course of doing so, institutions likely need to document the history of actions that have been taken, and ensure that the actions are just in specific and in aggregate. Auditing of the actions must occur by trusted experts.  The DSL framework provides the institution with a systematic (i.e. system-level) method for identifying issues of fiduciary responsibility, and documenting the methods for addressing the issues, both during and after implementation of the DSL.  Institutions can use several methods for this documentation as developed by others.  For example, institutions can publish for public consumption Declarations of Conformity (DoC) \cite{hind2018increasing} or factsheets \cite{european2020white}, giving stakeholders the opportunity to evaluate an institution's fiduciary responsibility.  Likewise, institutions can utilize contracts \cite{jacovi2021formalizing} for the benefit of stakeholders as well.  This particular aspect of the theory requires additional work beyond the scope of this article.

\section{Predictive Policing DSLs: Benefits, Risks, and Public Trust} \label{sec:case}


Over the past 14+ years, multiple urban police departments across the United States have sought and utilized algorithm-driven predictive policing technologies that evaluate massive volumes of historical crime/arrest data to predict high crime geographies/places or crime prone individuals, which help police leadership decide where and how to deploy officer resources. Predictive policing DSLs were pioneered by the Los Angeles Police Department (LAPD) in the 2000s and quickly spread to several major cities across the U.S. \cite{lau2020predictive}. There are two forms of predictive policing technology: place-based and person-based \cite{lau2020predictive}. Place-based predictive policing is the most extensively used method and leverages pre-existing crime data to identify places and times with a high probability of crime. Alternatively, person-based predictive policing looks for risk variables like previous arrests or victimization trends to identify individuals or groups who are likely to commit a crime or be a victim of one. 


Proponents of predictive policing technology assert that benefits of the technology include assisting police in forecasting crimes more objectively, precisely, and effectively than traditional policing methods and investigation techniques \cite{lau2020predictive,mohler2015randomized}. Predictive policing is intended as an automated tool to reduce primary reliance on officer instincts to forecast crime \cite{lau2020predictive}, thus increasing officer safety and the accuracy of crime prediction. Technology designers claim that they are not only capable of substantially reducing violent crimes such as murder, aggravated assault, and robbery, but also of removing bias from police decision-making \cite{lau2020predictive,thomson2018predictive}. However, claims that predictive policing reduces crime have been disputed \cite{santos2019predictive,mohler2015randomized} (see also \cite{hardyns2018predictive,meijer2019predictive}). 

The majority of risks associated with predictive policing technologies are related to the black box effect described in Section \ref{sec:dsl}. Predictive policing DSLs rely on previous crime data that are often incomplete due to a large percentage of crime being unknown or unreported, and/or racially biased due to the disparate arrests of African American and Hispanic people when compared to Whites across time \cite{carson2014prisoners,rosenberg2017comparing}. Scholarship has posited that racism and bias are systemically entrenched in the criminal justice system (CJS), facilitating the disproportionate mass incarceration of people of color, and influence police practices, policies, and behaviors on the ground \cite{alexander2011the,richardson2019dirty,ritchie2017invisible}. Black, Brown, and low-income communities have been over-policed historically due in part to the social acceptance of racism and high crime rates that are often related to poor structural conditions and a lack of access to resources associated with systemic inequities in the CJS and society at large \cite{alexander2011the}. Considering these arguments, the methodology through which officers collect the data impacting the pre-processing stage of the DSL may be influenced by systemic racism and racial bias in policing; and predictive policing algorithms (model building layer) rely on such data to generate place- or person-based predictions in the post-processing stages of the DSL \cite{richardson2019dirty}, which can perpetuate or reinforce historical prejudices in policing practices and policies \cite{aclu2016statement,barrett2017reasonably,edwards2016predictive,lau2020predictive,richardson2019dirty} \cite{oneil2016weapons,edwards2016predictive,lau2020predictive,richardson2019dirty}. Compounding the severity of such concerns, the manner in which police develop and administer predictive policing DSLs often lacks mechanisms to hold departments accountable for the interpretation of predictions and the decisions made/actions taken based on those predictions \cite{aclu2016statement,barrett2017reasonably,bhuiyan2021lapd,edwards2016predictive,lau2020predictive,moravec2019do,richardson2019dirty}. In the same vein, algorithmic predictions (DSL post-processing layer) can influence how police officers view the neighborhoods they are patrolling, and the ways in which they perceive individuals’ criminal propensity within those (primarily Black and Latino) communities \cite{ferguson2012predictive,tarantola2020predictive}, which ultimately affects probable cause for arrest decision-making.  

Scholars and activists have also posited that predictive policing DSLs facilitate increased and unjustified police stop, search, and seizure decision-making that can violate the Fourth Amendment of the U.S. constitution \cite{ferguson2012predictive,lau2020predictive}. Privacy concerns of predictive policing DSLs include eroding public anonymity through expanding webs of surveillance in the U.S., and creating networks of personal information that can be shared across police departments, accessed through illegal computer hacking or system breaches, or mishandled by officers \cite{ferguson2017rise}. Arguably, these outcomes negate any potential benefits of predictive policing technology \cite{richardson2019dirty}. Such risks have resulted in public (as well as data scientist) protests, boycotting, and privacy protection activism across the U.S. where predictive policing technology has been proposed or used \cite{aougab2020boycott,bhuiyan2021lapd,heaven2021predictive,hvistendahl2021how,linder2020why}.  


\subsection{The Black Box Effect: A Case Study of the Los Angeles Police Department (LAPD)} \label{ssec:black}

The LAPD began testing and implementing a person-based predictive policing DSL known as "Operation Laser" (referred to as, LASER--Los Angeles Strategic Extraction and Restoration) in 2011. LASER identified repeat offenders, and produced bulletins with their photos and physical descriptions so law enforcement could find/identify those individuals to prevent their future criminal activity \cite{moravec2019do}. Examining LASER's operation within the DSL framework described in this paper, the LASER algorithm utilized criminal history data to identify individuals most likely to commit a violent crime as part of the pre-processing layer of the DSL. ``Chronic offenders'' were ranked using a point system where factors like gang membership, number of perpetrated violent crimes, and interactions with officers were algorithmically assessed within the model building \& prediction stage (second DSL layer) of the DSL. Subsequently, those with higher numbers of ``points'' were placed on chronic offender bulletins within the "interpret" stage of DSL post-processing that were distributed to officers during the DSL post-processing "communicate" stage. These bulletins provided law enforcement with the identifying information necessary to specifically target (i.e. approach) those on the list as part of the "deploy" stage of the DSL post-processing layer.  

Andrew Ferguson, a law professor and nationally renowned predictive policing expert explained to CBS News \cite{baek2020lapd} that ``the LASER program was designed on the metaphor that they [the LAPD] were going to, like laser surgery, remove the tumors, the bad actors from the community…That idea, offensive as it is, was an idea of using some kind of algorithm to identify risk.'' Implementation of LASER led to public outcry specific to DSL predictions being used as a legal veneer for police brutality, mistreatment, and racial profiling against people of color \cite{aclu2016statement,ayers2008a,lapowski2018how}. The LAPD Inspector General conducted an internal audit of LASER in 2019 (eight years post-implementation), and found that Latinos/as and African Americans made up 84\% of ``active'' chronic offenders. 
The audit revealed numerous inconsistencies (relevant to the model building and interpret stages of the DSL) where the LASER algorithm identified and labeled individuals as ``chronic offenders'' \cite{baek2020lapd,smith2019review}. More specifically, 44\% of labeled ``chronic'' offenders had never been arrested or only had one arrest for some type of violent crime, and nearly 10\% had no ``quality interactions'' with law enforcement \cite{baek2020lapd,smith2019review}. The program was discontinued in 2019 following the audit. 

The LAPD also contracted with PredPol in 2011, which utilized historical property crime data to produce ``hot spot'' predictions with a high likelihood of vehicle theft and burglary \cite{lau2020predictive}. PredPol applies an ``earthquake'' crime prediction method which--like earthquakes and aftershocks--smaller crimes lead to bigger crimes and occur in near proximity to one another \cite{bhuiyan2021lapd,tonkin2011linking,yang2019predictive}. Like LASER, PredPol was in part intended to prevent subjective judgments and implicit bias as part of officer deployment decisions \cite{mooney2020is}. However, activists and anti-predictive policing community members maintain that the overwhelming bias of PredPol--i.e. hot spot map predictions identifying primarily Black and Brown neighborhoods--renders it unreliable and corrupt, and thus cannot be trusted and must be entirely dismantled \cite{bakke2018predictive,chang2018lapd,stop2018dismantling}. In relation to the post-processing layer of the DSL, when police leadership \textit{interpret} PredPol's problematic hot spot predictions and thereafter decide to assign higher or increasing numbers of officers (\textit{communicate} stage of the DSL) to patrol African American and Hispanic/Latino/a/x communities (\textit{deploy} stage of the DSL), the likelihood of civil rights and civil liberties violations increases. According to the recreancy theorem \cite{sapp2021public}, such actions within a black box do not reflect fiduciary responsibility, which further helps to explain Los Angeles residents' criticisms of, and wariness about, the department and its use of predictive policing. We envision an idealized DSL in which a (human) policing/data specialist receives the output of the model and interprets this output within a broader context. This interpretation is communicated to someone with the authority to implement policing strategies who decides how to utilize the information and then determines the deployment of resources. 

Anti-predictive policing protests, advocacy organization mobilization, and academic criticism began to escalate in 2016, and 17 groups, including the American Civil Liberties Union (ACLU) and the NAACP, signed a widely circulated statement indicating their concerns about the reinforcement of racial bias associated with predictive policing technologies \cite{aclu2016statement} and the lack of transparency about development and use of the DSLs from the institutions that administer them \cite{lau2020predictive}. PredPol was discontinued by the LAPD in 2020; the department claimed that this action was taken because of COVID-19 financial constraints \cite{lau2020predictive,miller2020lapd}. 

\subsection{Manifesting Public Mistrust in Predictive Policing}

Referring back to Section \ref{sec:dsl}, depending on the nature of the predictive policing prediction output, police personnel interpret (i.e. evaluate) the prediction, communicate the evaluation with other personnel, and ultimately deploy (i.e. take action) police resources accordingly. Fiduciary responsibility manifested in the DSL framework does much to explain public mistrust of predictive policing as implemented by the LAPD. Public mistrust was largely associated with the LAPD’s lack of transparency as the department developed and administered the DSLs across the pre-processing, data modeling, and post-processing layers of the DSL (see \cite{lau2020predictive}). The LAPD began utilizing LASER and PredPol as mechanisms of crime control in Los Angeles neighborhoods absent disclosures to the public about the basics, intricacies, development and administration of, or decision-making associated with, the technologies \cite{moravec2019do}, which can be explained by the black box effect depicted in Figure \ref{fig:2-blackbox} rather than in Figure \ref{fig:1-blackbox}. \footnote{It is important to note that this argument is formulated based on the authors' search of media and government press release databases.}  

More specifically, the minimal information that the LAPD did share with the public lacked thorough and transparent details on the multiple layers across the DSLs, such as the types of data utilized (i.e. pre-processing layer), how the algorithm(s)/technologies were developed and produced predictions (i.e. model building layer), and how policing decisions were made and actions taken based on those predictions (i.e. post-processing layer). A search of Newsbank's Access World News database revealed that news stories about LAPD's predictive policing technologies began to show up sporadically in the media and in LAPD press releases from 2012-2014 (e.g. see \cite{bailey2012stopping,lapd2013lapd,madrigal2013toward}) (2-3 years post-implementation), which may be why protests and activism against the technologies only began to gain traction around 2016 (5-years post-implementation). An organization known as, ``Stop LAPD Spying Coalition'' requested public records about the LAPD’s use of LASER in 2017 and 2018 due to concerns about the unfair LAPD targeting of, and forced interactions with, Los Angeles residents, and then filed a lawsuit against the department in 2018 when the information was not provided \cite{baek2020lapd}. As a result, the LAPD began to release many of the records to the public, albeit slowly \cite{baek2020lapd}. 

The public is unlikely to support departments administering predictive policing technologies when they fail to provide transparent, thorough, and honest information on their benefits and risks across the end stages of the DSL. The black box effect during the post-processing stages of the LAPD's DSLs (i.e. decision-making/actions taken) fueled public mistrust specific to system predictions providing a covert excuse for racialized enforcement of the law and institutional racism \cite{mooney2020is,richardson2019dirty}. Predictive policing DSLs can enable ‘‘tech-washing’’ where communities and people of color are specifically, disproportionately, and (potentially) unjustly targeted with a façade of data-driven ethics and objectivity \cite{egbert2021discrimination}. Through tech-washing, departments can operate absent sanctions or responsibility, which can reinforce harmful stereotypes and systemic injustice and facilitate their perpetuation. In this manner, the LAPD failed to meet its fiduciary responsibility to ensure that the DSL was not misused by police.

\subsection{Facilitating Fiduciary Responsibility: Open-box DSLs} 

The LAPD's deployment of LASER and PredPol illustrates how a societal institution’s failure to meet fiduciary responsibility \cite{sapp2021public} in its development and implementation of DSLs resulted in the public perceiving the department as lacking integrity--untrustworthy, dishonest, racist, and unjust toward people and communities of color.  Department actions taken following DSL prediction output in the post-processing stages--hidden from stakeholders--are suspected to be immoral, discriminatory, and harmful to communities (e.g. disproportionate targeting and arrests of people of color, overpolicing of their communities, and worsening mass incarceration).  We argue that police departments’ development and deployment of predictive policing DSLs in secret lacks accountability and has great potential to upset the functioning of society \cite{sapp2021public}. 
The LAPD’s failure to build trust in their implementation of the predictive policing DSLs resulted in residents losing trust in, and support for, not only those DSLs but the LAPD overall. 

Such outcomes demonstrate the necessity of assessing and potentially altering the design, deployment, and usage of a DSL within the pre-processing, model building, and post-processing layers. Specifically, these outcomes show that DSLs must operate within an open-box regime, as we depict in Figure \ref{fig:1-blackbox}. This regime provides the mechanisms for stakeholders to be confident that the institution is meeting its fiduciary responsibility through understanding interpretations of model outputs and thorough communication of the DSL's ultimate decisions.  Additionally, the open-box regime facilitates stakeholder and expert auditing of the overall DSL.

Here are some specific examples of open-box DSL actions police departments could take as part of the last two stages of the post-processing layer to enhance fiduciary responsibility. In the Communicate stage for example, police leadership could provide frequent and descriptive/transparent press releases and social media posts that indicate the need for predictive policing, the data being used, how community members will be affected, process and status of design and implementation, and methods used to reduce bias. Similarly, departments could also hold media events or town halls to answer questions from the press and public, as well as take note of any concerns that should be considered prior to deploying resources/actions (i.e., decision-making) in impacted communities (i.e., last stage of the post-processing layer). In regards to the latter, police leadership should consider again communicating with the public about how such concerns were or will be addressed. Utilizing multiple methods of communication may have a stronger effect on public perceptions of the department's integrity and fiduciary responsibility.

As part of the Deploy stage, law enforcement may want to use a document similar to a contract (or create mandatory procedural guidelines) that explicates the actions they will take based on the output/prediction of the predictive policing DSL.  This document could be updated periodically to reflect lessons learned (and subsequently alter or enhance communication content and strategies) and public feedback received across the duration of AI-based decision-making and deployment. We do not provide suggestions here for the Interpret stage of the DSL post-processing layer because this stage is internal to the police department and dependent on insider (i.e., police officer/leadership) knowledge.





\section{Conclusion and Future Work}

Facilitating public trust in AI, and those institutions that deploy them, is imperative for maintaining social cohesion.  The recreancy theorem delineates three dimensions of public support for technical innovations. In this paper, we considered the dimension of fiduciary responsibility, which is the public's perception that a technology is designed and employed with integrity and honesty, performs as intended, does not create disparate impacts to vulnerable populations, and is not misused by agents. To formalize AI fiduciary responsibility, we described AI within the larger perspective of a DSL. The DSL framework provides multiple methods to precisely describe fiduciary responsibility of technologies affecting the public welfare and how institutions can meet their fiduciary responsibility. We investigated the example of the LAPD not meeting its fiduciary responsibility in its deployment of predictive policing technology.

We envision future work in at least two directions. First, the description of fiduciary responsibility within the DSL framework can be further refined and quantified. We have introduced multiple aspects and manifestations of fiduciary responsibility within DSLs, but did not consider system dynamics in our investigation--this consideration warrants a full, separate analysis given the complexity of dynamics involving DSLs. In addition, our work here was only qualitative in nature, and we did not propose mechanisms for institutions to quantify whether they have met their fiduciary responsibilities.  Second, further development of the DSL framework more broadly can be utilized in a number of potential ways to facilitate public trust in AI through documentation, regulatory requirements, and a technologically-aware public. This, too, warrants a full, separate analysis from the one we have presented in this work. 

The current paper uniquely situated sources/causes of mistrust in varying DSL contexts wherein each layer/stage has processes through which fiduciary responsibility can (or cannot) be addressed. We illustrated the importance (and necessity) of embedding fiduciary responsibility across the DSL workflow over time wherein institutions' decisions and deployment of actions in the post-processing layer can effect the public in profound and consequential ways.

\section*{Acknowledgment}  Shannon B. Harper and Eric S. Weber were supported by the National Science Foundation and the National Geospatial Intelligence Agency under award \#1830254.  Eric S. Weber was also supported by the National Science Foundation under award \#1934884. 

\bibliographystyle{IEEEtran}
\bibliography{bibs/dspipeline,bibs/harper,bibs/trust,bibs/sapp}

\begin{thebibliography}{100}
\providecommand{\url}[1]{#1}
\csname url@samestyle\endcsname
\providecommand{\newblock}{\relax}
\providecommand{\bibinfo}[2]{#2}
\providecommand{\BIBentrySTDinterwordspacing}{\spaceskip=0pt\relax}
\providecommand{\BIBentryALTinterwordstretchfactor}{4}
\providecommand{\BIBentryALTinterwordspacing}{\spaceskip=\fontdimen2\font plus
\BIBentryALTinterwordstretchfactor\fontdimen3\font minus
  \fontdimen4\font\relax}
\providecommand{\BIBforeignlanguage}[2]{{%
\expandafter\ifx\csname l@#1\endcsname\relax
\typeout{** WARNING: IEEEtran.bst: No hyphenation pattern has been}%
\typeout{** loaded for the language `#1'. Using the pattern for}%
\typeout{** the default language instead.}%
\else
\language=\csname l@#1\endcsname
\fi
#2}}
\providecommand{\BIBdecl}{\relax}
\BIBdecl

\bibitem{sapp2021public}
S.~G. Sapp, S.~Dorius, K.~Bertelson, and S.~Harper, ``Public support for
  government use of network surveillance: An empirical assessment of public
  understanding of ethics in science administration,'' \emph{Public
  Understanding of Science}, 2021.

\bibitem{kao2022effect}
Y.~Kao and S.~G. Sapp, ``The effect of cultural values and institutional trust
  on public perceptions of government use of network surveillance,'' 2022,
  technology in Society (in press).

\bibitem{sapp2009consumer}
S.~G. Sapp, C.~Arnot, J.~Fallon, T.~Fleck, D.~Soorholtz, M.~Sutton-Vermeulen,
  and J.~J.~H. Wilson, ``Consumer trust in the u.s. food system: An examination
  of the recreancy theorem,'' \emph{Rural Sociology}, vol.~74, pp. 525--545,
  2009.

\bibitem{mohler2015randomized}
G.~O. Mohler, M.~B. Short, S.~Malinowski, M.~Johnson, G.~E. Tita, A.~L.
  Bertozzi, and P.~J. Brantingham, ``Randomized controlled field trials of
  predictive policing,'' \emph{Journal of the American statistical
  association}, vol. 110, no. 512, pp. 1399--1411, 2015.

\bibitem{lum2016predict}
\BIBentryALTinterwordspacing
K.~Lum and W.~Isaac, ``To predict and serve?'' \emph{Significance}, vol.~13,
  no.~5, pp. 14--19, 2016. [Online]. Available:
  \url{https://rss.onlinelibrary.wiley.com/doi/abs/10.1111/j.1740-9713.2016.00960.x}
\BIBentrySTDinterwordspacing

\bibitem{ensign2018runaway}
D.~Ensign, S.~A. Friedler, S.~Neville, C.~Scheidegger, and
  S.~Venkatasubramanian, ``Runaway feedback loops in predictive policing,'' in
  \emph{Conference on Fairness, Accountability and Transparency}, PMLR.\hskip
  1em plus 0.5em minus 0.4em\relax New York, NY USA: ACM, 2018, pp. 160--171.

\bibitem{hengstler2016applied}
M.~Hengstler, E.~Enkel, and S.~Duelli, ``Applied artificial intelligence and
  trust—the case of autonomous vehicles and medical assistance devices,''
  \emph{Technological Forecasting and Social Change}, vol. 105, pp. 105--120,
  2016.

\bibitem{siau2018building}
K.~Siau and W.~Wang, ``Building trust in artificial intelligence, machine
  learning, and robotics,'' \emph{Cutter business technology journal}, vol.~31,
  no.~2, pp. 47--53, 2018.

\bibitem{dobbe2018broader}
R.~Dobbe, S.~Dean, T.~Gilbert, and N.~Kohli. (2018) A broader view on bias in
  automated decision-making: Reflecting on epistemology and dynamics.

\bibitem{lee2004trust}
J.~D. Lee and K.~A. See, ``Trust in automation: Designing for appropriate
  reliance,'' \emph{Human factors}, vol.~46, no.~1, pp. 50--80, 2004.

\bibitem{JS16}
L.~Jaume-Palasí and M.~Spielkamp, ``Ethics and algorithmic processes for
  decision making and decision support.'' 2016, algorithm Watch, Working Paper
  \#2.

\bibitem{MATWF}
B.~D. Mittelstadt, P.~Allo, M.~Taddeo, S.~Wachter, and L.~Floridi, ``The ethics
  of algorithms: Mapping the debate,'' \emph{Big Data \& Society}, vol.~3,
  no.~2, p. 2053951716679679, 2016.

\bibitem{BS15}
S.~Barocas and A.~D. Selbst, ``Big data's disparate impact,'' 2015, sSRN
  Scholarly Paper, Rochester, NY: Social Science Research Network. At:
  http://papers.ssrn.com/abstract=2477899.

\bibitem{mouzannar2019fair}
H.~Mouzannar, M.~I. Ohannessian, and N.~Srebro, ``From fair decision making to
  social equality,'' in \emph{Proceedings of the Conference on Fairness,
  Accountability, and Transparency}.\hskip 1em plus 0.5em minus 0.4em\relax New
  York, NY, USA: ACM, 2019, pp. 359--368.

\bibitem{knowles2021sanction}
B.~Knowles and J.~T. Richards, ``The sanction of authority: Promoting public
  trust in {AI},'' in \emph{Proceedings of the 2021 ACM Conference on Fairness,
  Accountability, and Transparency}.\hskip 1em plus 0.5em minus 0.4em\relax New
  York, NY, USA: ACM, 2021, pp. 262--271.

\bibitem{biswas2021art}
S.~Biswas, M.~Wardat, and H.~Rajan, ``The art and practice of data science
  pipelines: A comprehensive study of data science pipelines in theory,
  in-the-small, and in-the-large,'' in \emph{ICSE'2022: The 44th International
  Conference on Software Engineering}.\hskip 1em plus 0.5em minus 0.4em\relax
  New York, NY, USA: ACM, May 21-May 29 2022.

\bibitem{nguyen2019machine}
G.~Nguyen, S.~Dlugolinsky, M.~Bob{\'a}k, V.~Tran, {\'A}.~L. Garc{\'\i}a,
  I.~Heredia, P.~Mal{\'\i}k, and L.~Hluch{\`y}, ``Machine learning and deep
  learning frameworks and libraries for large-scale data mining: a survey,''
  \emph{Artificial Intelligence Review}, vol.~52, no.~1, pp. 77--124, 2019.

\bibitem{olson2016evaluation}
R.~S. Olson, N.~Bartley, R.~J. Urbanowicz, and J.~H. Moore, ``Evaluation of a
  tree-based pipeline optimization tool for automating data science,'' in
  \emph{Proceedings of the genetic and evolutionary computation conference
  2016}.\hskip 1em plus 0.5em minus 0.4em\relax New York, NY, USA: ACM, 2016,
  pp. 485--492.

\bibitem{wickham2019data}
\BIBentryALTinterwordspacing
H.~Wickham. (2014) Data science: how is it different to statistics? IMS
  Bulletin. [Online]. Available:
  \url{https://imstat.org/2014/09/04/data-science-how-is-it-different-to-statistics\%E2\%80\%89/}
\BIBentrySTDinterwordspacing

\bibitem{hong2017build}
\BIBentryALTinterwordspacing
S.~A. Hong and T.~Hunter. (2017) Build, scale, and deploy deep learning
  pipelines with ease. Databricks.com. [Online]. Available:
  \url{https://databricks.com/blog/2017/09/06/build-scale-deploy-deep-learning-pipelines-ease.html}
\BIBentrySTDinterwordspacing

\bibitem{todd2017computing}
S.~Todd and D.~Dietrich, ``Computing resource re-provisioning during data
  analytic lifecycle,'' 2017, uS Patent 9,619,550.

\bibitem{garcia2018context}
R.~Garcia, V.~Sreekanti, N.~Yadwadkar, D.~Crankshaw, J.~E. Gonzalez, and J.~M.
  Hellerstein, ``Context: The missing piece in the machine learning
  lifecycle,'' in \emph{KDD CMI Workshop}, vol. 114.\hskip 1em plus 0.5em minus
  0.4em\relax New York, NY, USA: ACM, 2018.

\bibitem{polyzotis2018data}
N.~Polyzotis, S.~Roy, S.~E. Whang, and M.~Zinkevich, ``Data lifecycle
  challenges in production machine learning: a survey,'' \emph{ACM SIGMOD
  Record}, vol.~47, no.~2, pp. 17--28, 2018.

\bibitem{zhou2019how}
\BIBentryALTinterwordspacing
L.~Zhou. (2019) How to build a better machine learning pipeline. Datanami.com.
  [Online]. Available:
  \url{https://www.datanami.com/2018/09/05/how-to-build-a-better-machine-learning-pipeline/}
\BIBentrySTDinterwordspacing

\bibitem{wing2019data}
\BIBentryALTinterwordspacing
J.~M. Wing. (2019) The data life cycle. Harvard Data Science Review. [Online].
  Available: \url{https://hdsr.mitpress.mit.edu/pub/577rq08d/release/3}
\BIBentrySTDinterwordspacing

\bibitem{ashmore2021assuring}
R.~Ashmore, R.~Calinescu, and C.~Paterson, ``Assuring the machine learning
  lifecycle: Desiderata, methods, and challenges,'' \emph{ACM Computing Surveys
  (CSUR)}, vol.~54, no.~5, pp. 1--39, 2021.

\bibitem{freudenburg1993risk}
W.~R. Freudenburg, ``Risk and recreancy: Weber, the division of labor, and the
  rationality of risk perceptions,'' \emph{Social forces}, vol.~71, no.~4, pp.
  909--932, 1993.

\bibitem{alario2003paradoxes}
M.~Alario and W.~Freudenburg, ``The paradoxes of modernity: scientific
  advances, environmental problems, and risks to the social fabric?''
  \emph{Sociological Forum}, vol.~18, no.~2, pp. 193--214, 2003.

\bibitem{roth1978economy}
G.~Roth and C.~Wittich, \emph{Economy and Society: an outline of interpretive
  sociology}.\hskip 1em plus 0.5em minus 0.4em\relax Berkeley, USA: University
  of California Press, 1978.

\bibitem{colquitt2007trust}
J.~A. Colquitt, B.~A. Scott, and J.~A. LePine, ``Trust, trustworthiness, and
  trust propensity: a meta-analytic test of their unique relationships with
  risk taking and job performance.'' \emph{Journal of applied psychology},
  vol.~92, no.~4, p. 909, 2007.

\bibitem{ball2019institutional}
K.~Ball, S.~Degli~Esposti, S.~Dibb, V.~Pavone, and E.~Santiago-Gomez,
  ``Institutional trustworthiness and national security governance: evidence
  from six european countries,'' \emph{Governance}, vol.~32, no.~1, pp.
  103--121, 2019.

\bibitem{barber1983logic}
B.~Barber, \emph{The logic and limits of trust}.\hskip 1em plus 0.5em minus
  0.4em\relax New Brunswick, USA: Rutgers University Press, 1983.

\bibitem{blomqvist2005trust}
K.~Blomqvist, ``Trust in a dynamic environment-fast trust as a threshold
  condition for asymmetric technology partnership formation in the ict
  sector,'' \emph{Trust in Pressure: Investigations of trust and trust building
  in uncertain circumstances}, pp. 127--147, 2005.

\bibitem{blomqvist2005playing}
K.~Blomqvist, P.~Hurmelinna, and R.~Sepp{\"a}nen, ``Playing the collaboration
  game right—balancing trust and contracting,'' \emph{Technovation}, vol.~25,
  no.~5, pp. 497--504, 2005.

\bibitem{deutsch1958trust}
M.~Deutsch, ``Trust and suspicion,'' \emph{Journal of conflict resolution},
  vol.~2, no.~4, pp. 265--279, 1958.

\bibitem{sapp-dm2009consumer}
S.~G. Sapp and T.~Downing-Matibag, ``Consumer acceptance of food irradiation: a
  test of the recreancy theorem,'' \emph{International Journal of Consumer
  Studies}, vol.~33, no.~4, pp. 417--424, 2009.

\bibitem{kim2012participation}
S.~Kim and J.~Lee, ``E-participation, transparency, and trust in local
  government,'' \emph{Public administration review}, vol.~72, no.~6, pp.
  819--828, 2012.

\bibitem{hardin2001conceptions}
R.~Hardin, ``Conceptions and explanations of trust.'' in \emph{Trust in
  Society}, K.~Cook, Ed.\hskip 1em plus 0.5em minus 0.4em\relax Russell Sage
  Foundation, 2001, pp. 3--39.

\bibitem{earle1995social}
T.~C. Earle and G.~Cvetkovich, \emph{Social trust: Toward a cosmopolitan
  society}.\hskip 1em plus 0.5em minus 0.4em\relax Westport, USA: Greenwood
  Publishing Group, 1995.

\bibitem{siegrist2000perception}
M.~Siegrist and G.~Cvetkovich, ``Perception of hazards: The role of social
  trust and knowledge,'' \emph{Risk analysis}, vol.~20, no.~5, pp. 713--720,
  2000.

\bibitem{cooper2008importance}
C.~A. Cooper, H.~G. Knotts, and K.~M. Brennan, ``The importance of trust in
  government for public administration: The case of zoning,'' \emph{Public
  Administration Review}, vol.~68, no.~3, pp. 459--468, 2008.

\bibitem{schoorman2007integrative}
F.~D. Schoorman, R.~C. Mayer, and J.~H. Davis, ``An integrative model of
  organizational trust: Past, present, and future,'' pp. 344--354, 2007.

\bibitem{sapp2013science}
S.~G. Sapp, P.~F. Korsching, C.~Arnot, and J.~J.~H. Wilson, ``Science
  communication and the rationality of public opinion formation,''
  \emph{Science Communication}, vol.~35, pp. 734--757, 2013.

\bibitem{anderson2018artificial}
J.~Anderson, L.~Rainie, and A.~Luchsinger, ``Artificial intelligence and the
  future of humans,'' \emph{Pew Research Center}, vol.~10, p.~12, 2018.

\bibitem{hitlin2019facebook}
\BIBentryALTinterwordspacing
P.~Hitlin and L.~Rainie. (2019) Facebook algorithms and personal data. Pew
  Research Center. [Online]. Available:
  \url{https://www.pewresearch.org/internet/2019/01/16/facebook-algorithms-and-personal-data/}
\BIBentrySTDinterwordspacing

\bibitem{olmstead2017americans}
K.~Olmstead and A.~Smith, ``Americans and cybersecurity,'' \emph{Pew Research
  Center}, vol.~26, pp. 311--327, 2017.

\bibitem{rainie2017code}
\BIBentryALTinterwordspacing
L.~Rainie, J.~Anderson, and D.~Page. (2017) Code-dependent: Pros and cons of
  the algorithm age. Pew Research Center. [Online]. Available:
  \url{https://www.pewresearch.org/internet/2017/02/08/code-dependent-pros-and-cons-of-the-algorithm-age/}
\BIBentrySTDinterwordspacing

\bibitem{rainie2017internet}
\BIBentryALTinterwordspacing
L.~Rainie and J.~Anderson. (2017) The internet of things connectivity binge:
  what are the implications? Pew Research Center. [Online]. Available:
  \url{https://www.pewresearch.org/internet/2017/06/06/the-internet-of-things-connectivity-binge-what-are-the-implications/}
\BIBentrySTDinterwordspacing

\bibitem{lewis1985trust}
J.~D. Lewis and A.~Weigert, ``Trust as a social reality,'' \emph{Social
  forces}, vol.~63, no.~4, pp. 967--985, 1985.

\bibitem{hofstede2006intrinsic}
G.~J. Hofstede, ``Intrinsic and enforceable trust: a research agenda,''
  European Association of Agricultural Economists, 99th Seminar, February 8-10,
  2006, Tech. Rep., 2006, bonn, Germany.

\bibitem{WHKA14}
M.~Wisner, S.~Hammer, E.~Kurdyukova, and E.~Andre, ``Trust-based
  decision-making for the adaptation of public displays in changing social
  contexts,'' 2014, journal of Trust Management 20: 41-46.

\bibitem{ibm2020trusting}
\BIBentryALTinterwordspacing
{IBM Research AI}. (2020) Trusting {AI}. IBM. [Online]. Available:
  \url{https://www.research.ibm.com/artificial-intelligence/trusted-ai/}
\BIBentrySTDinterwordspacing

\bibitem{european2020white}
\BIBentryALTinterwordspacing
{European Commission}. (2020) White paper: On artificial intelligence--a
  european approach to excellence and trust. European Commission. [Online].
  Available:
  \url{https://ec.europa.eu/info/sites/info/files/commission-white-paper-artificial-intelligence-feb2020\_en.pdf}
\BIBentrySTDinterwordspacing

\bibitem{YH08}
Z.~Yan and S.~Holtmanns, ``Trust modeling and management: from social trust to
  digital trust,'' 2008, iGI Global, Hershey.

\bibitem{harper2014trust}
R.~Harper, \emph{Trust, computing, and society}.\hskip 1em plus 0.5em minus
  0.4em\relax Cambridge, UK: Cambridge University Press, 2014.

\bibitem{jobin2019global}
A.~Jobin, M.~Ienca, and E.~Vayena, ``The global landscape of {AI} ethics
  guidelines,'' \emph{Nature Machine Intelligence}, vol.~1, no.~9, pp.
  389--399, 2019.

\bibitem{toreini2020relationship}
E.~Toreini, M.~Aitken, K.~Coopamootoo, K.~Elliott, C.~G. Zelaya, and
  A.~Van~Moorsel, ``The relationship between trust in {AI} and trustworthy
  machine learning technologies,'' in \emph{Proceedings of the 2020 conference
  on fairness, accountability, and transparency}.\hskip 1em plus 0.5em minus
  0.4em\relax New York, NY, USA: ACM, 2020, pp. 272--283.

\bibitem{arnold2019factsheets}
M.~Arnold, R.~K. Bellamy, M.~Hind, S.~Houde, S.~Mehta, A.~Mojsilovi{\'c},
  R.~Nair, K.~N. Ramamurthy, A.~Olteanu, D.~Piorkowski \emph{et~al.},
  ``Factsheets: Increasing trust in {AI} services through supplier's
  declarations of conformity,'' \emph{IBM Journal of Research and Development},
  vol.~63, no. 4/5, pp. 6--1, 2019.

\bibitem{jacovi2021formalizing}
A.~Jacovi, A.~Marasovi{\'c}, T.~Miller, and Y.~Goldberg, ``Formalizing trust in
  artificial intelligence: Prerequisites, causes and goals of human trust in
  {AI},'' in \emph{Proceedings of the 2021 ACM Conference on Fairness,
  Accountability, and Transparency}.\hskip 1em plus 0.5em minus 0.4em\relax New
  York, NY, USA: ACM, 2021, pp. 624--635.

\bibitem{mayer1995integrative}
R.~C. Mayer, J.~H. Davis, and F.~D. Schoorman, ``An integrative model of
  organizational trust,'' \emph{Academy of management review}, vol.~20, no.~3,
  pp. 709--734, 1995.

\bibitem{hawley2014trust}
K.~Hawley, ``Trust, distrust and commitment,'' \emph{No{\^u}s}, vol.~48, no.~1,
  pp. 1--20, 2014.

\bibitem{ribeiro2016should}
M.~T. Ribeiro, S.~Singh, and C.~Guestrin, ````why should {I} trust you?''
  {E}xplaining the predictions of any classifier,'' in \emph{Proceedings of the
  22nd ACM SIGKDD international conference on knowledge discovery and data
  mining}.\hskip 1em plus 0.5em minus 0.4em\relax New York, NY, USA: ACM, 2016,
  pp. 1135--1144.

\bibitem{mittelstadt2019explaining}
B.~Mittelstadt, C.~Russell, and S.~Wachter, ``Explaining explanations in
  {AI},'' in \emph{Proceedings of the conference on fairness, accountability,
  and transparency}.\hskip 1em plus 0.5em minus 0.4em\relax New York, NY, USA:
  ACM, 2019, pp. 279--288.

\bibitem{richards2020methodology}
J.~Richards, D.~Piorkowski, M.~Hind, S.~Houde, and A.~Mojsilovi{\'c}. (2020) A
  methodology for creating {AI} factsheets. IBM Research.

\bibitem{hind2018increasing}
M.~Hind, S.~Mehta, A.~Mojsilovic, R.~Nair, K.~N. Ramamurthy, A.~Olteanu, and
  K.~R. Varshney, ``Increasing trust in {AI} services through supplier’s
  declarations of conformity,'' 2018, arXiv preprint arXiv:1808.07261.

\bibitem{davis1989perceived}
F.~D. Davis, ``Perceived usefulness, perceived ease of use, and user acceptance
  of information technology,'' \emph{MIS quarterly}, pp. 319--340, 1989.

\bibitem{davis1993user}
------, ``User acceptance of information technology: system characteristics,
  user perceptions and behavioral impacts,'' \emph{International journal of
  man-machine studies}, vol.~38, no.~3, pp. 475--487, 1993.

\bibitem{wachter2017counterfactual}
S.~Wachter, B.~Mittelstadt, and C.~Russell, ``Counterfactual explanations
  without opening the black box: Automated decisions and the {GDPR},''
  \emph{Harv. JL \& Tech.}, vol.~31, p. 841, 2017.

\bibitem{gilpin2022explanation}
\BIBentryALTinterwordspacing
L.~Gilpin, A.~Paley, M.~Alam, S.~Spurlock, and K.~Hammond, ````explanation'' is
  not a technical term: {T}he problem of ambiguity in {XAI},'' 2022. [Online].
  Available: \url{https://arxiv.org/abs/2207.00007}
\BIBentrySTDinterwordspacing

\bibitem{kohli2018translation}
\BIBentryALTinterwordspacing
N.~Kohli, R.~Barreto, and J.~A. Kroll, ``Translation tutorial: A shared lexicon
  for research and practice in human-centered software systems,'' in \emph{1st
  Conference on Fairness, Accountability, and Transparency}, New York, NY,
  2018, p.~7. [Online]. Available:
  \url{https://facctconference.org/static/tutorials/fatconf18\_lexicon\_tutorial.pdf}
\BIBentrySTDinterwordspacing

\bibitem{lau2020predictive}
\BIBentryALTinterwordspacing
T.~Lau. (2020) Predictive policing explained. Brennan Center. [Online].
  Available:
  \url{https://www.brennancenter.org/our-work/research-reports/predictive-policing-explained}
\BIBentrySTDinterwordspacing

\bibitem{thomson2018predictive}
\BIBentryALTinterwordspacing
S.~Thomson. (2018) 'predictive policing': Law enforcement revolution or just
  new spin on old biases? depends who you ask. CBC News. [Online]. Available:
  \url{https://www.cbc.ca/news/world/crime-los-angeles-predictive-policing-algorithms-1.4826030}
\BIBentrySTDinterwordspacing

\bibitem{santos2019predictive}
R.~B. Santos, \emph{Predictive policing: Where’s the evidence?}\hskip 1em
  plus 0.5em minus 0.4em\relax Cambridge, UK: Cambridge University Press, 2019,
  pp. 366--396.

\bibitem{hardyns2018predictive}
W.~Hardyns and A.~Rummens, ``Predictive policing as a new tool for law
  enforcement? recent developments and challenges,'' \emph{European journal on
  criminal policy and research}, vol.~24, no.~3, pp. 201--218, 2018.

\bibitem{meijer2019predictive}
A.~Meijer and M.~Wessels, ``Predictive policing: Review of benefits and
  drawbacks,'' \emph{International Journal of Public Administration}, vol.~42,
  no.~12, pp. 1031--1039, 2019.

\bibitem{carson2014prisoners}
\BIBentryALTinterwordspacing
E.~A. Carson and D.~Golinelli. (2014) Prisoners in 2012: Trends in admissions
  and releases, 1991–2012. US Department of Justice. [Online]. Available:
  \url{https://bjs.ojp.gov/content/pub/pdf/p12tar9112.pdf}
\BIBentrySTDinterwordspacing

\bibitem{rosenberg2017comparing}
A.~Rosenberg, A.~K. Groves, and K.~M. Blankenship, ``Comparing black and white
  drug offenders: Implications for racial disparities in criminal justice and
  reentry policy and programming,'' \emph{Journal of drug issues}, vol.~47,
  no.~1, pp. 132--142, 2017.

\bibitem{alexander2011the}
M.~Alexander, ``The new {J}im {C}row,'' \emph{Ohio St. J. Crim. L.}, vol.~9,
  p.~7, 2011.

\bibitem{richardson2019dirty}
R.~Richardson, J.~M. Schultz, and K.~Crawford, ``Dirty data, bad predictions:
  How civil rights violations impact police data, predictive policing systems,
  and justice,'' \emph{NYUL Rev. Online}, vol.~94, p.~15, 2019.

\bibitem{ritchie2017invisible}
A.~J. Ritchie, \emph{Invisible no more: Police violence against Black women and
  women of color}.\hskip 1em plus 0.5em minus 0.4em\relax Beacon press, 2017.

\bibitem{aclu2016statement}
\BIBentryALTinterwordspacing
A.~C.~L. Union. (2016) Statement of concern about predictive policing by aclu
  and 16 civil rights privacy, racial justice, and technology organizations.
  American Civil Liberties Union. [Online]. Available:
  \url{https://www.aclu.org/other/statement-concern-about-predictive-policing-aclu-and-16-civil-rights-privacy-racial-justice}
\BIBentrySTDinterwordspacing

\bibitem{barrett2017reasonably}
L.~Barrett, ``Reasonably suspicious algorithms: predictive policing at the
  united states border,'' \emph{NYU Rev. L. \& Soc. Change}, vol.~41, p. 327,
  2017.

\bibitem{edwards2016predictive}
A.~Edwards, ``Big data, predictive machines and security: The minority report
  1,'' in \emph{The Routledge handbook of technology, crime and justice}.\hskip
  1em plus 0.5em minus 0.4em\relax Oxfordshire, UK: Routledge, 2017, pp.
  451--461.

\bibitem{oneil2016weapons}
C.~O'Neil, \emph{Weapons of Math Destruction: How Big Data Increases Inequality
  and Threatens Democracy}.\hskip 1em plus 0.5em minus 0.4em\relax New York,
  USA: Crown Books, 2016.

\bibitem{bhuiyan2021lapd}
\BIBentryALTinterwordspacing
J.~Bhuiyan. (2021) {LAPD} ended predictive policing programs amid public
  outcry. {A} new effort shares many of their flaws. The Guardian. [Online].
  Available:
  \url{https://www.theguardian.com/us-news/2021/nov/07/lapd-predictive-policing-surveillance-reform}
\BIBentrySTDinterwordspacing

\bibitem{moravec2019do}
\BIBentryALTinterwordspacing
E.~R. Moravec. (2019) Do algorithms have a place in policing? The Atlantic.
  [Online]. Available:
  \url{https://www.theatlantic.com/politics/archive/2019/09/do-algorithms-have-place-policing/596851/}
\BIBentrySTDinterwordspacing

\bibitem{ferguson2012predictive}
A.~G. Ferguson, ``Predictive policing and reasonable suspicion’(2012),''
  \emph{Emory Law Journal}, vol.~62, p. 259, 2012, available at SSRN:
  https://papers.ssrn.com/sol3/papers.cfm?abstract\_id=2050001.

\bibitem{tarantola2020predictive}
\BIBentryALTinterwordspacing
A.~Tarantola. (2020) 'predictive policing' could amplify today's law
  enforcement issues. Engadget. [Online]. Available:
  \url{https://www.engadget.com/predictive-policing-privacy-civil-rights-dangers-133040971.html}
\BIBentrySTDinterwordspacing

\bibitem{ferguson2017rise}
A.~G. Ferguson, \emph{The rise of big data policing}.\hskip 1em plus 0.5em
  minus 0.4em\relax New York, NY, USA: New York University Press, 2017.

\bibitem{aougab2020boycott}
\BIBentryALTinterwordspacing
T.~Aougab and et. al., ``Boycott collaboration with police,'' \emph{Notices
  Amer. Math. Soc.}, vol.~67, no.~9, p. 1293, 2020, date accessed: 12-02-2021.
  [Online]. Available:
  \url{https://www.ams.org/journals/notices/202009/rnoti-p1293.pdf}
\BIBentrySTDinterwordspacing

\bibitem{heaven2021predictive}
\BIBentryALTinterwordspacing
W.~D. Heaven. (2020) Predictive policing algorithms are racist. {T}hey need to
  be dismantled. MIT Technology Review. [Online]. Available:
  \url{https://www.technologyreview.com/2020/07/17/1005396/predictive-policing-algorithms-racist-dismantled-machine-learning-bias-criminal-justice/}
\BIBentrySTDinterwordspacing

\bibitem{hvistendahl2021how}
\BIBentryALTinterwordspacing
M.~Hvistendahl. (2021) How the {LAPD} and {P}alantir use data to justify racist
  policing. The Intercept. [Online]. Available:
  \url{https://theintercept.com/2021/01/30/lapd-palantir-data-driven-policing/}
\BIBentrySTDinterwordspacing

\bibitem{linder2020why}
\BIBentryALTinterwordspacing
C.~Linder. (2020) Why hundreds of mathematicians are boycotting predictive
  policing. Popular Mechanics. [Online]. Available:
  \url{https://www.popularmechanics.com/science/math/a32957375/mathematicians-boycott-predictive-policing/}
\BIBentrySTDinterwordspacing

\bibitem{baek2020lapd}
\BIBentryALTinterwordspacing
G.~Baek and T.~Mooney. (2020) {LAPD} not giving up on data-driven policing,
  even after scrapping controversial program. CBS News. [Online]. Available:
  \url{https://www.cbsnews.com/news/los-angeles-police-department-laser-data-driven-policing-racial-profiling-2-0-cbsn-originals-documentary/}
\BIBentrySTDinterwordspacing

\bibitem{ayers2008a}
\BIBentryALTinterwordspacing
I.~Ayres and J.~Borowsky. (2008) A study of racially disparate outcomes in the
  {L}os {A}ngeles {P}olice {D}epartment. ACLU of Southern California. [Online].
  Available:
  \url{https://www.aclusocal.org/sites/default/files/wp-content/uploads/2015/09/11837125-LAPD-Racial-Profiling-Report-ACLU.pdf}
\BIBentrySTDinterwordspacing

\bibitem{lapowski2018how}
\BIBentryALTinterwordspacing
I.~Lapowski. (2018) How the {LAPD} uses data to predict crime. wired.com.
  [Online]. Available:
  \url{https://www.wired.com/story/los-angeles-police-department-predictive-policing/}
\BIBentrySTDinterwordspacing

\bibitem{smith2019review}
\BIBentryALTinterwordspacing
M.~P. Smith. (2019) Review of selected {L}os {A}ngeles {P}olice {D}epartment
  data-driven policing strategies. Los Angeles Police Commission. [Online].
  Available:
  \url{https://a27e0481-a3d0-44b8-8142-1376cfbb6e32.filesusr.com/ugd/b2dd23\_21f6fe20f1b84c179abf440d4c049219.pdf}
\BIBentrySTDinterwordspacing

\bibitem{tonkin2011linking}
M.~Tonkin, J.~Woodhams, R.~Bull, J.~W. Bond, and E.~J. Palmer, ``Linking
  different types of crime using geographical and temporal proximity,''
  \emph{Criminal Justice and Behavior}, vol.~38, no.~11, pp. 1069--1088, 2011.

\bibitem{yang2019predictive}
F.~Yang, ``Predictive policing,'' in \emph{Oxford Research Encyclopedia of
  Criminology and Criminal Justice}.\hskip 1em plus 0.5em minus 0.4em\relax
  Oxford, UK: Oxford University Press, 2019.

\bibitem{mooney2020is}
\BIBentryALTinterwordspacing
T.~Mooney and G.~Baek. (2020) Is artificial intelligence making racial
  profiling worse? CBS News. [Online]. Available:
  \url{https://www.cbsnews.com/news/artificial-intelligence-racial-profiling-2-0-cbsn-originals-documentary/}
\BIBentrySTDinterwordspacing

\bibitem{bakke2018predictive}
E.~Bakke, ``Predictive policing: the argument for public transparency,''
  \emph{NYU Ann. Surv. Am. L.}, vol.~74, p. 131, 2018.

\bibitem{chang2018lapd}
\BIBentryALTinterwordspacing
C.~Chang. (2018) {LAPD} officials defend predictive policing as some groups
  call for its end. Lexipol. [Online]. Available:
  \url{https://www.police1.com/patrol-issues/articles/lapd-officials-defend-predictive-policing-as-some-groups-call-for-its-end-PNfxLd2b6JajAZDs/}
\BIBentrySTDinterwordspacing

\bibitem{stop2018dismantling}
\BIBentryALTinterwordspacing
S.~L.~S. Coalition. (2018) Before the bullet hits the body: Dismantling
  predictive policing in los angeles. Stop LAPD Spying Coalition. [Online].
  Available:
  \url{https://stoplapdspying.org/wp-content/uploads/2018/05/Before-the-Bullet-Hits-the-Body-Report-Summary.pdf}
\BIBentrySTDinterwordspacing

\bibitem{miller2020lapd}
\BIBentryALTinterwordspacing
L.~Miller. (2020) {LAPD} will end controversial program that aimed to predict
  where crimes would occur. Los Angeles Times. [Online]. Available:
  \url{https://www.latimes.com/california/story/2020-04-21/lapd-ends-predictive-policing-program}
\BIBentrySTDinterwordspacing

\bibitem{bailey2012stopping}
\BIBentryALTinterwordspacing
R.~Bailey. (2012) Stopping crime before it starts. Reason. [Online]. Available:
  \url{https://reason.com/2012/07/10/predictive-policing-criminals-crime/}
\BIBentrySTDinterwordspacing

\bibitem{lapd2013lapd}
\BIBentryALTinterwordspacing
L.~A.~P. Department. (2013) {LAPD} foothill community police station
  announces” international predpol day of action”. Los Angeles Police
  Department. [Online]. Available:
  \url{https://www.lapdonline.org/newsroom/lapd-foothill-community-police-station-announces-international-predpol-day-of-action-na13136bb/}
\BIBentrySTDinterwordspacing

\bibitem{madrigal2013toward}
\BIBentryALTinterwordspacing
A.~C. Madrigal. (2013) Toward a complex, realistic, and moral tech criticism.
  The Atlantic. [Online]. Available:
  \url{https://www.theatlantic.com/technology/archive/2013/03/toward-a-complex-realistic-and-moral-tech-criticism/273996/}
\BIBentrySTDinterwordspacing

\bibitem{egbert2021discrimination}
S.~Egbert and M.~Mann, ``Discrimination in predictive policing: The (dangerous)
  myth of impartiality and the need for sts analysis,'' in \emph{Automating
  Crime Prevention, Surveillance, and Military Operations}.\hskip 1em plus
  0.5em minus 0.4em\relax Springer, 2021, pp. 25--46.

\end{thebibliography}


\begin{IEEEbiography}[{\includegraphics[width=1in,height=1.25in,clip,keepaspectratio]{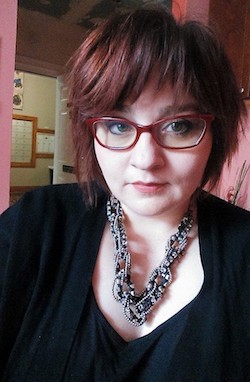}}]
{Shannon B. Harper} is an Assistant Professor of criminal justice in the Department of Sociology and Criminal Justice at Iowa State University. Dr. Harper’s research explores the relationship between intimate partner violence and homicide, as well as how crime victims perceive the usefulness and accessibility of institutions that provide victim services, including the criminal justice system. Such scholarship includes focus on police decision-making processes that influence their interactions with victims and the public at large. Her work is published in numerous highly ranked criminological/sociological journals, including the \textit{Journal of Interpersonal Violence}, \textit{Public Understanding of Science}, \textit{Feminist Criminology}, and the \textit{American Journal of Criminal Justice}.
\end{IEEEbiography}

\begin{IEEEbiography}[{\includegraphics[width=1in,height=1.25in,clip,keepaspectratio]{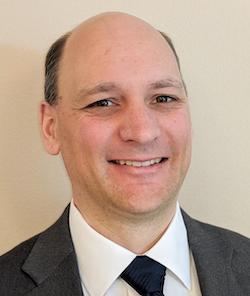}}]
{Eric S. Weber} is Professor and Chair of Mathematics at Iowa State University.  Dr. Weber holds a Ph.D. in Mathematics from the University of Colorado.  His research interests include harmonic analysis and approximation theory.  Past research includes developing novel wavelet transforms for image processing and reproducing kernel methods for the harmonic analysis of fractals.  Current research projects include the development of new algorithms for processing distributed spatiotemporal datasets in order to increase understanding of human dynamics; extending alternating projection methods for optimization in non-Euclidean geometries; using harmonic analysis techniques for understanding the approximation properties of neural networks; and developing machine learning techniques to improve the diagnosis of severe wind occurrences.

\end{IEEEbiography}

\end{document}